\documentclass{elsart}
\usepackage[dvips]{epsfig}

\begin{document}

\begin{frontmatter}

\title{Signature of exotic particles in light by light scattering} 
\author{G. Tavares--Velasco \thanksref{gtv}}  
\address{Departamento de F\'{\i}sica, CINVESTAV, Apartado Postal
14--740, 07000, M\'exico D. F., M\'exico} 
\author{J.J. Toscano \thanksref{jjt}}
\address{Facultad de Ciencias F\'{\i}sico Matem\'aticas, Benem\'erita 
Universidad Aut\'onoma de Puebla,
Apartado Postal 1152, Puebla, Pue., M\'exico.} 
\thanks [gtv] {E--mail: gtv@fis.cinvestav.mx}
\thanks [jjt] {E--mail: jtoscano@fcfm.buap.mx}

\maketitle 

\begin{abstract} 

We discuss the implications on light by light scattering of two kind of
exotic particles: doubly charged scalar bosons and doubly charged
fermions; the virtual effects of a nonstandard singly charged gauge
boson are also examined. These particles, if their masses lie in
the range 0.1--1.0 TeV, will have a clear signature in the future linear
colliders. The present analysis has the advantage that it depends only
on electromagnetic symmetry, so it is applicable to any model which
predicts this class of particles. In particular, our results have
interesting consequences on left-right models and their supersymmetric
extension. 

\end{abstract} 

\end{frontmatter}

%pacs numbers
%\pacs{14.70.Bh, 14.80.Cp, 14.70.Pw, 12.60.Jv}

The major dream in particle physics is a final theory of elementary
interactions. The standard model (SM) leaves many unaddressed questions,
so it is only one step toward the achievement of such a theory. In the
attempt to take one step forward many extensions have been conjectured,
resulting in the prediction of new particles, whose experimental evidence
would point in the right direction. In this letter we will examine the
implications of different kinds of exotic particles on light by light
scattering \cite{photon}--\cite{Jikia}, which has been proposed recently
as an useful mode to detect virtual effects of new physics at the future
linear colliders (LC) \cite{Renard}. Our study includes doubly charged
scalar bosons and fermions, as well as a singly charged gauge boson
heavier than the SM one. The mass range studied will be 0.1--1 TeV, which
would be at the reach of LC \cite{LinColl}. Since the $\gamma \gamma$
scattering amplitude is proportional to the electric charge factor $Q^4$,
the contribution of particles with charge greater than the unity, in terms
of the positron charge, would enhance dramatically the respective cross
section, resulting in a distinctive signal of new physics. Moreover, the
structure of the cross section is entirely dictated by the spin of the
particles circulating in the loop and by electromagnetic symmetry, thus it
is not necessary to make further assumptions about a specific model. As a
consequence, our analysis is applicable to any model which predicts such
particles. Nevertheless, the main motivation of our work resides in two
popular extensions of the standard model, namely left-right symmetric
models (LRM)  \cite{Mohapatra} and their supersymmetric extension (SUSYLR) 
\cite{Susylrm}, where these exotic particles are a natural prediction and
they might provide the most distinctive signature of these models.  We
will begin by presenting a brief outline of these extensions.

Doubly charged Higgs bosons emerge in many extensions of the SM Higgs
sector \cite{Higgs}. Unacceptable effects on the electroweak tree level
relation $\rho \equiv m_W^2/(\cos \theta_W m_Z^2) \approx 1$ may arise
from exotic representations which have triplets or higher
representations with a neutral member, but it is possible to overcome
this problem by recoursing to extra assumptions. This is the case in a
popular class of models with both doublets and triplets, where a
custodial $\mathrm{SU(2)}$ symmetry is invoked to protect the $\rho$
relation \cite{Georgi}. In general, the $\rho$ problem is avoided in
representations without neutral members or either in models where the
vacuum expectation value (VEV) of the neutral component of the Higgs
multiplets vanishes. More complicated possibilities arise in extensions
of the $\mathrm{SU(2)}\times \mathrm{U(1)}$ gauge group. For instance,
the minimal version of LRM, which are based on the $\mathrm{SU(2)_L}
\times \mathrm{SU(2)_R} \times \mathrm{U(1)_{B-L}}$ gauge group, requires
the presence of one bidoublet as well as left and right triplets. These
models predict the existence of new physics at an intermediate scale
provided by the VEV of the neutral component of the right triplet,
parity-symmetry would be restored at this scale. The VEV's of the
neutral components of the bidoublet are identified with the Fermi scale.
As far as the left triplet is concerned, it is only required to preserve
left-right symmetry, and its neutral component can get a VEV constrained
to be small to maintain the $\rho$ relation. In its minimal realization,
LRM predict fourteen physical Higgs bosons but we are only interested on
the doubly charged bosons as it is likely that they give a clear
signature of new physics. In the gauge sector, the only new contribution
to light by light scattering comes from a charged right--handed gauge
boson. The low-energy implications of LRM depend strongly on the
structure and the vacuum stability of the Higgs potential. It was shown
that a careful analysis of the most general Higgs potential consistent
with the low-energy data might not lead to new physics at a low scale
\cite{Gunion}. However, if new physics featured by discrete symmetries
is imposed, the most delicate terms of the Higgs potential can be
eliminated, allowing the existence of a scale that would be accessible
at LC. In particular, in these scenarios some Higgs bosons might be
light, with masses of the order of the Fermi scale. 

With regard to doubly charged fermions, they appear first in natural
lepton models \cite{Wilc}. More recently, they have emerged naturally
from the supersymmetric extension of LRM. At this respect, there is the
belief that supersymmetry (SUSY), and in particular its minimal
low-energy realization, the supersymmetric standard model (MSSM), is a
natural candidate to supersede the SM. Although MSSM offers solution to
problems not explained by the SM it has many undesirable features,
namely it predicts large $CP$ violating effects, it allows the presence
of baryon and lepton number violating terms in the lagrangian, and it
forbids the existence of massive neutrinos if global $R$--parity
is to be conserved. These problems may be cured by considering the
supersymmetric extension of LRM, SUSYLR \cite{Mohapatra2}, which has
also the attractiveness of the presence of a low-energy scale $m \sim
M_R^2/M_{\mathrm{Planck}}$, where $M_R$ is the scale of left-right
symmetry breaking. It happens that some singly and doubly charged Higgs
scalars and the respective superpartners have their masses proportional
to $m$. Very recently it has been argued that two interesting
possibilities arise depending if the vacuum of the theory does or does
not conserve $R$--parity \cite{Mohapatra3}. When the vacuum conserves
$R$--parity, low energy data set a lower limit on $M_R$ of about
$10^{10}$ GeV. On the other hand, in the scenario where the vacuum state
breaks $R$--parity spontaneously, it exists an upper limit of about 10
TeV for $M_R$. It follows that even in the case of $M_R$ in the range
$10^{10}-10^{12}$ GeV, there is the possibility of some light doubly
charged Higgs bosons and Higgsinos. This is an important motivation to
study the implications of doubly charged scalars and fermions at LC
through light by light scattering. 

The main virtue of light by light scattering is that to a certain extent
it is a model-independent process: the structure of its amplitude is
entirely dictated by the spin of the virtual particles as well as
electromagnetic symmetry, and of course by the number of such particles.
Therefore, the only dependence on a specific model is given by the mass
and electric charge of its particle content. As was pointed out in
\cite{Renard}, in the SM the helicity amplitudes of the $\gamma \gamma$
scattering are almost purely imaginary at high energies, precisely in
the range where there is the possibility of observing the appearance of
fields associated with models beyond the SM. There follows that if the
contribution arising from additional charged particles has an
appreciable imaginary part, the virtual effects would become evident
through the interference with the SM contribution even if the respective
contribution is too small to be detected by itself. Another interesting
feature of light by light scattering is that the amplitude arising from
loops with the same particle circulating on them turns out to be
proportional to the charge factor $Q^4$, the consequence is that the
cross section gets enhanced dramatically when the contribution of a
doubly charged particle is considered. With all these properties, the
process $\gamma \gamma \to \gamma \gamma$ provides an excellent
mechanism to search for virtual effects of new physics at LC: the
signature of a certain particle with a particular spin and charge
depends only on its mass, and it is not necessary to consider
model-dependent parameters or make further assumptions as it does the
case when direct production is studied. 

We will proceed to discuss our results. We have considered the
implications on light by light scattering of doubly charged scalars and
fermions with masses in the range 0.1--0.5 TeV. As far as the singly
charged gauge boson is concerned, we have studied the case in which its
mass is greater that the existing bound of $550$ GeV, obtained if it is
assumed a light right-handed neutrino \cite{Caso}. For the purpose of this
work, it is sufficient to analyze unpolarized cross sections. A more
detailed study will be presented elsewhere \cite{Toscano}, including
polarized cross sections and the implications of the technical details of
LC, together with the study of exotic particles not discussed in here:
doubly charged gauge bosons and exotic quarks with charges $5/3~e$ and
$-4/3~e$, which are predicted by some $\mathrm{SU(3)_c} \times
\mathrm{SU(3)_L} \times \mathrm{U(1)_N}$ models \cite{Pisano}. The
helicity amplitudes of contributions of loops with scalars, fermions and
gauge bosons are well known \cite{Karplus}--\cite{Renard}. To obtain the
cross section we have worked with the exact amplitudes without making any
simplification but to consider the observable cross section, the
integration has been constrained to $|\cos\theta| \leq 30^{\mathrm{o}}$.
The numerical analysis was done with the program {\small{FF}}
\cite{Oldenborgh}.

The case where singly charged fermions and scalar bosons are involved
was studied with details in \cite{Renard}, in the context of SUSY models
\footnote{We do not show these results, but it must be noticed that we
find a nice agreement with \cite{Renard}. }. The remarkably properties
of the $\gamma \gamma$ scattering acquire new dimensions when particles
with a charge greater than the unity, in units of the positron charge,
are involved. This is shown through Figs. 1-6, where we have plotted
separately the contributions of each exotic particle for different
values of its mass. In Fig. 1, it is shown the respective contribution
as well as the interference between a doubly charged fermion and the SM
contribution, scaled by the SM unpolarized cross section
$\sigma_{\mathrm{SM}}$. It must be noticed that, though the virtual
effects arise predominantly from the interference term, in the case of a
light fermion with a mass of about 100 GeV even its own contribution
plays an important role. This seems to contradict the well known result
that at energies about 300 GeV the top contribution is negligible with
respect to the $W_L$ term. The explanation resides in the powerful
charge factor: as the helicity amplitudes for the contribution of a
certain particle are proportional to the factor $Q^4$, the cross section
arising from a doubly charged fermion turns out to be improved by the
factor $2^8/(3 (2/3)^4)^2=3^6$ in comparison with the case of an
up--type quark, whereas the interference term gets enhanced by the
factor $3^3$. When the mass $M_{\hat{\delta}_{++}}$ of the doubly
charged fermion is greater than 200 GeV, its contribution to the
unpolarized cross section tends to be suppressed with respect to the
interference term. It is also interesting to note that the virtual
effects are always observed above the threshold $\sqrt{s} \geq 2
M_{\hat{\delta}_{++}}$. This fact is explained because of the character
predominantly imaginary of the SM helicity amplitudes at energies above
300 GeV, at the same time the new contributions are purely real below
the threshold and complex above it. As the interference is given by $2
\Re(A_{\mathrm{SM}} A_{\mathrm{New}})$, it follows that the virtual
effects will become evident above the threshold. To realize the
magnitude of the deviation from the SM, we have plotted in Fig. 2 the
effect that would be observed if in addition to the SM particle content,
a doubly charged fermion is included. It is clear that the signature of
such an exotic particle would be very distinctive. The sensitivity of
the unpolarized cross section at a linear collider running at energies
in the range of 350--800 GeV has been examined for the case of a singly
charged fermion \cite{Renard}, it was found that for a chargino with a
mass of 100-250 GeV the signal varies between 3 SD--1 SD. It is evident
that in the case of a doubly charged fermion we should expect an
important increment \cite{Toscano}. Although the signature of a doubly
charged scalar is not as spectacular as that of a fermion with the same
charge, the situation is also promising as it is depicted in Figs 3--4.
In Fig. 3 it can be seen that the virtual effects come predominantly
from the interference with the SM particles, the result is reflected on
the deviation from the unpolarized cross section $\sigma_{\mathrm{SM}}$
(Fig. 4). If we compare this particular case with that of a singly
charged scalar with about the same mass, we find that, since the
dominant effect comes from interference, the deviation for a doubly
charged scalar is larger for a factor of $2^4$. As a result, the
possibility of observing the virtual effects of a relatively heavy
doubly charged scalar would be increased as compared to the case of the
singly charged scalar. Finally, for completeness we have studied the
potential effects of a relatively light singly charged gauge boson. The
motivation is that it has been noticed that the existence of such a
particle would have important implications to elucidate some aspects of
SUSYLR \cite{Mohapatra3}. We have plotted in Figs 5--6 the deviation
from the SM cross section as caused by a singly charged gauge boson with
a mass in the range 0.55--1 TeV. It can be seen that, as expected, the
enhancement of the cross section is not as important as the ones arising
from doubly charged particles. However, at energies of about 2 TeV, the
signal would be more important than the one coming from a doubly charged
scalar. In contrast to the situation of fermions and scalars, where the
deviation from the SM cross section is most important near the
threshold, the one coming from a gauge boson is larger far beyond.

Our results show that the signature of exotic particles would be
distinctive enough in $\gamma \gamma$ scattering to provide evidence of
new physics. In particular, in the context of SUSYLR, an interesting
implication is that a doubly charged Higgs scalar boson or a doubly
charged Higgsino with masses in the range 100-500 GeV would produce a
clear signature. This signal would be more distinctive at LC that the
one which could provide a chargino or a sfermion with the same or even
with a lighter mass. In addition, the existence of several doubly
charged particles will enhance spectacularly the cross section and as a
result particle counting might be possible through light by light
scattering. If SUSYLR is realized in nature, then it would exist the
possibility that low--energy remnant doubly charged Higgs bosons and
doubly charged Higgsinos would exist, as it has been suggested recently
in conjecturing some scenarios \cite{Mohapatra3}. If this possibility
became true, it is likely that this kind of particles would be
discovered by direct production \cite{Cuypers} by the time that LC would
be a reality. In this situation, $\gamma \gamma$ scattering might be an
effective process to probe the theory with a high precision.

In conclusion, due to its outstanding properties, light by light
scattering rises as an invaluable process to search indirectly signals
of doubly charged particles at the planned linear colliders. The most
remarkably feature is that the signal arising from a certain particle
depends only on its mass. As an alternative to direct production, where
model--dependent parameters have to be considered, light by light
scattering might offer the possibility of testing to a great detail the
properties of new charged particles. In particular, it would aid also to
elucidate some characteristics of a specific model, for instance the
number of particles of a certain kind.\\ 

\ack{We acknowledge support from CONACYT and SNI (M\'exico).}

\onecolumn
\newpage 

\begin{figure}[p] 
\begin{center} 
\epsfig{file=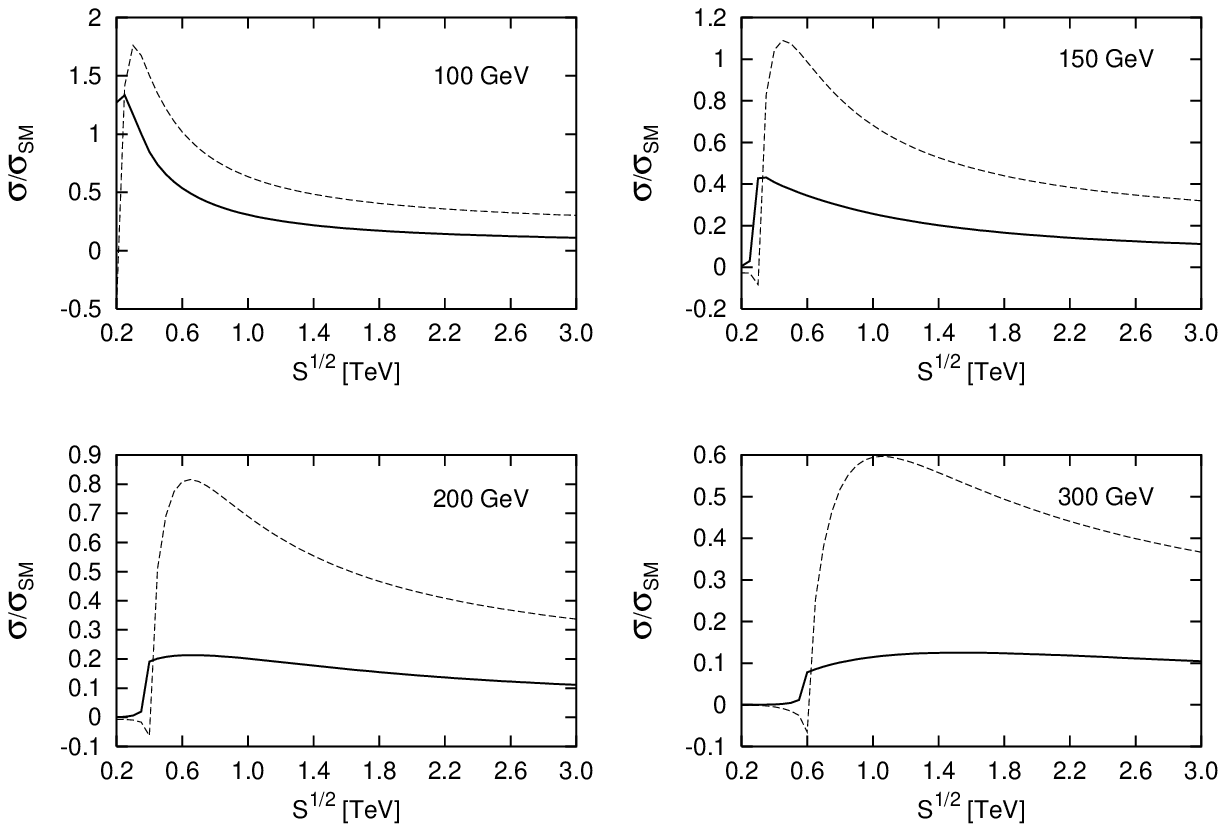,width=5in}
\caption{Interference (dashes) between the SM amplitude and that of a
doubly charged fermion $\hat{\delta}^{++}$ as well as its contribution
(solid line) to the unpolarized cross section, for different values of
the fermion mass. The values are scaled by the SM unpolarized cross
section.} 
\end{center} 
\end{figure} 

\begin{figure}[p] 
\begin{center} 
\epsfig{file=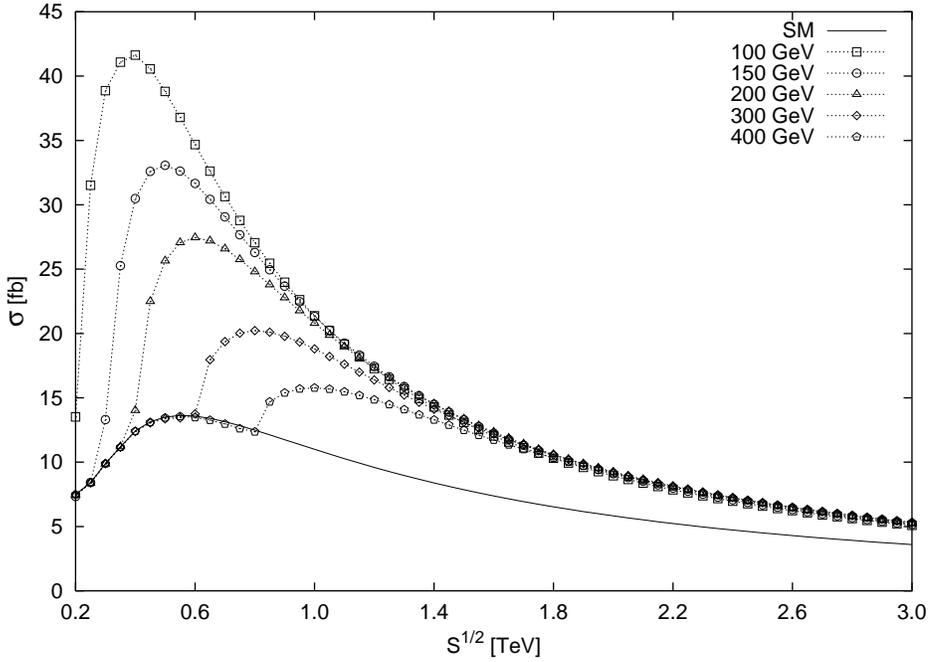,width=5in}
\caption{Deviaton from the SM unpolarized cross section when a doubly
charged fermion $\hat{\delta}^{++}$ contributes, for different values of
the fermion mass.} 
\end{center} 
\end{figure} 

\begin{figure}[p] 
\begin{center} 
\epsfig{file=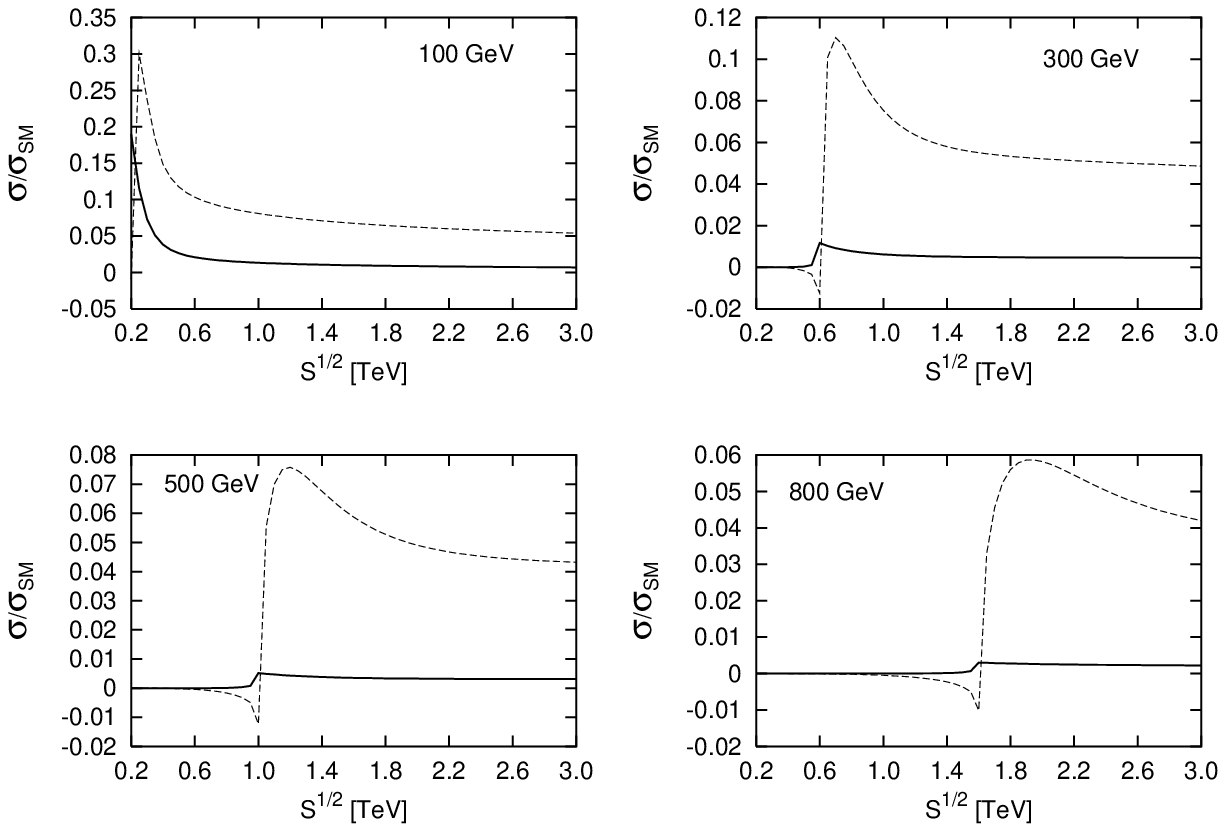,width=5in}
\caption{The same as in Fig. 1 when a doubly charged scalar
$\delta^{++}$ is involved.} 
\end{center} 
\end{figure} 

\begin{figure}[p] 
\begin{center} 
\epsfig{file=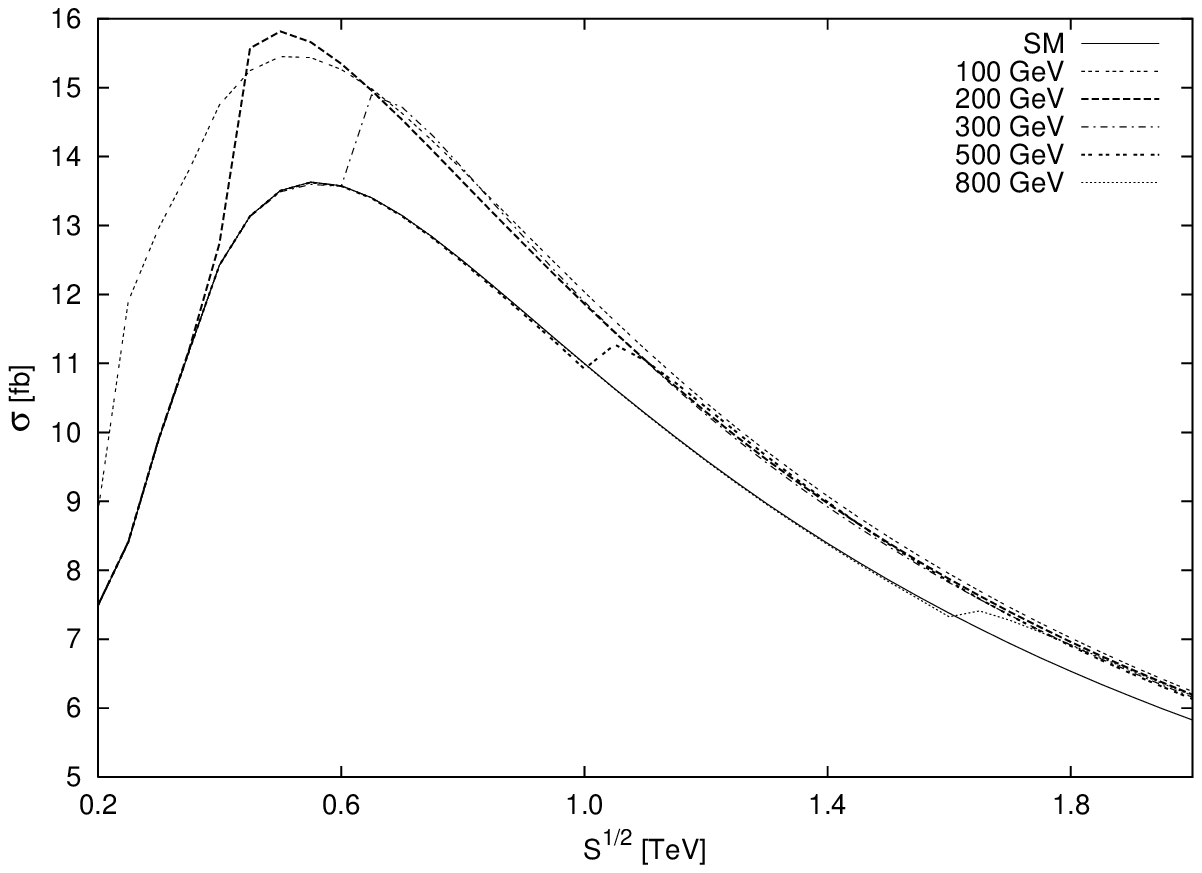,width=5in}
\caption{The same as in Fig. 2 when a doubly charged scalar
$\delta^{++}$ is involved.} 
\end{center} 
\end{figure} 

\begin{figure}[p] 
\begin{center} 
\epsfig{file=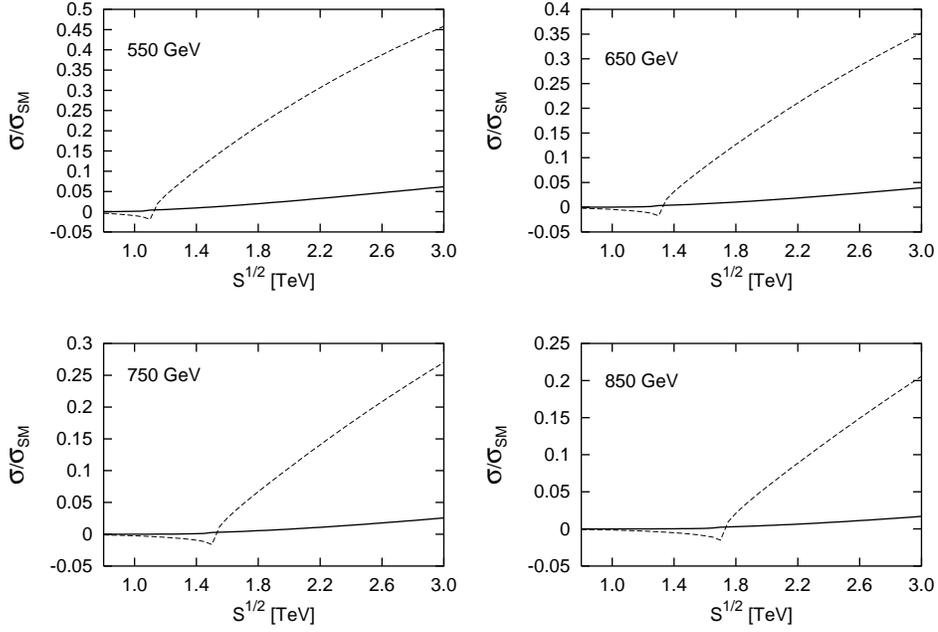,width=5in}
\caption{The respective plot as in Fig. 1 when a singly charged gauge
boson $W_R$ is involved.} 
\end{center} 
\end{figure} 

\begin{figure}[p] 
\begin{center} 
\epsfig{file=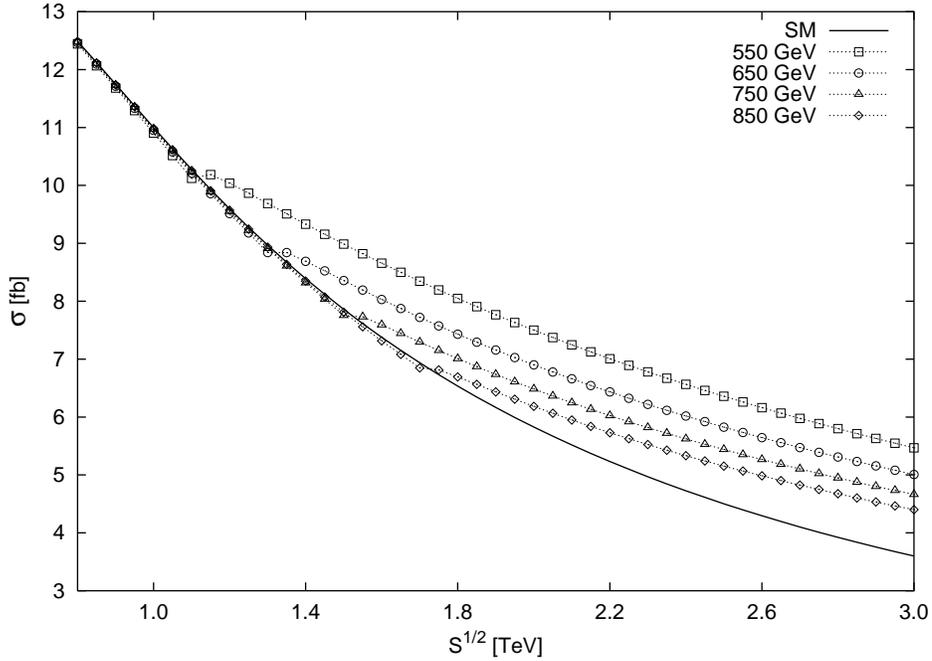,width=5in}
\caption{The same as in Fig. 2 when a singly charged gauge boson $W_R$
is involved.} 
\end{center} 
\end{figure} 

\end{document}